%% file: fin2.tex
\documentclass[aps,prl,nofootinbib,10pt,tightenlines,twocolumn,floatfix,superscriptaddress,longbibliography]{revtex4-2}

\usepackage[utf8]{inputenc}          % file encoding
\usepackage[usenames,dvipsnames]{xcolor}

\usepackage{amsmath,amssymb,amsfonts} % Math symbols
\usepackage[linktocpage,breaklinks]{hyperref}
\usepackage{mathrsfs}
\usepackage{txfonts}
\usepackage{bm}
\usepackage{orcidlink}
\usepackage[normalem]{ulem}

\usepackage[scr=esstix]{mathalfa} 

\usepackage{cleveref}
\usepackage{comment}
\usepackage[maxfloats=256]{morefloats}

\hypersetup{colorlinks=true,
            citecolor=NavyBlue,
            linkcolor=NavyBlue,
            urlcolor=NavyBlue}

\newcommand{\phiN}{\phi_{\rm N}}
\newcommand{\phiNhat}{\hat{\phi}_{\rm N}}

\newcommand{\Lie}{\mathscr{L}}
\newcommand{\calP}{\mathscr{P}}
\newcommand{\rmd}{\mathrm{d}}

\newcommand{ \Sm }{ S_{\rm  matt} }
\newcommand{ \Sf }{ S_{\rm  field} }

\DeclareMathOperator*{\sym}{\rm sym}

\def\be{\begin{equation}}
\def\ee{\end{equation}}
\def\bes{\begin{subequations}}
\def\ees{\end{subequations}}
\def\bea{\begin{eqnarray}}
\def\eea{\end{eqnarray}}

\begin{document}

\title{Nonlinearly self-interacting extended bodies move as test bodies in effective external fields}

\author{Abraham I. Harte\orcidlink{0000-0002-5893-5680}}
\affiliation{Centre for Astrophysics and Relativity (CfAR),
	School of Mathematical Sciences, \\
	Dublin City University, Glasnevin, Dublin 9, Ireland}

\author{Francisco M. Blanco\orcidlink{0000-0002-7711-8395}}
\affiliation{Department of Physics, Cornell University, Ithaca, NY 14853, USA.}

\author{Éanna É. Flanagan\orcidlink{0000-0001-6818-3550}}
\affiliation{Department of Physics, Cornell University, Ithaca, NY 14853, USA.}

\begin{abstract}

In electromagnetism, linearized general relativity, and other contexts, previous work has shown that the laws of motion which govern compact, self-interacting bodies can be obtained by applying ``Detweiler-Whiting prescriptions'' to the laws of motion which govern test bodies. These prescriptions replace any field which appears in a test-body law of motion with a certain effective field which is a quasilocal functional of the physical variables -- a functional that can be interpreted as a regularization procedure in a point-particle limit. We generalize these results, presenting a formalism which allows Detweiler-Whiting prescriptions to be be directly derived for extended bodies, even in nonlinear field theories. If a generating functional with particular properties can be constructed, we find effective linear and angular momenta which evolve via Mathisson-Papapetrou-Dixon equations involving appropriate effective fields. These equations implicitly incorporate all self-force, self-torque, and extended-body effects. Although our main focus is on bodies coupled to nonlinear scalar fields, we also remark on the gravitational case. 
\end{abstract}

\maketitle

\vspace{0.1cm}

How do objects fall in gravitational fields? Given a Newtonian potential $\phiN$, one answer is that the center of mass $z$ of a small test body accelerates via $\ddot{z}_i = a_i^{\rm test}[\phiN] \equiv - \nabla_i \phiN(z)$. However, the test-body assumption which is needed to justify this expression is unreasonably strong. If an isolated, irregularly-shaped object with density $\rho$ is initially at rest, $a_i^{\rm test}[\phiN]$ predicts that the object's self-field displaces it by its own diameter within a timescale of order $(G \rho)^{-1/2}$. This evaluates to approximately 30 minutes even for a pebble, which is clearly erroneous.

Better-behaved equations of motion are obtained by showing, before any approximations are applied, that no net force can be exerted by a body's self-field $\phiN^{\rm self}$. Given any object with mass $m$, cancellations of internal ``action-reaction pairs'' show that in terms of the exact, integral expression
\begin{equation}
    a^{\rm exact}_i [\phiN] \equiv - \frac{1}{m} \int \rmd^3 x  \, \rho \nabla_i \phiN
    \label{accelN}
\end{equation}
for the acceleration, $a^{\rm exact}_i [\phiN] = a^{\rm exact}_i [\phiN-\phiN^{\rm self}]$. It is therefore irrelevant whether $\phiN$ or the external potential $\phiNhat \equiv \phiN-\phiN^{\rm self}$ is used in \textit{exact} expressions for the acceleration. However, these  expressions are difficult to apply in practice. It is typical to instead approximate accelerations using truncated multipolar expansions, and in that context, the monopolar $a_i^{\rm test} [\phiNhat]$ is often a much better \textit{approximation} than $a_i^{\rm test} [\phiN]$. This is because $\phiNhat$ excludes the small-scale structure which can make $\nabla_i \phiN(z)$ a poor approximation for the density-weighted average of $\nabla_i \phiN$ which appears in $a_i^{\rm exact}[\phiN]$.

What is important here is that a useful law of motion for self-interacting bodies can be obtained simply by applying the replacement $\phiN \to \phiNhat$ to a law which holds with no self-interaction at all. Identical replacement rules hold not only for accelerations, but also for torques, and they extend through all multipolar orders in Newtonian gravity \cite{Dix79, Damour300yrs, Harte_2015}. In fact, similar prescriptions hold even in relativistic theories, where they are sometimes referred to as Detweiler-Whiting prescriptions.

Detweiler and Whiting \cite{Detweiler:2002mi} found that for a point particle with scalar charge $q$ which is coupled to a massless Klein-Gordon field, there is an effective field $\hat{\phi}$ such that the force (including the self-force) is $- q \nabla_a \hat{\phi}$. Similarly, electrically-charged particles were found to accelerate via a Lorentz force law involving an effective electromagnetic field.   Uncharged particles in linearized general relativity were also shown to move on geodesics of an effective metric. These results all have the appearance of test-body laws of motion where physical fields have been replaced by effective fields. They preserve legitimate self-force effects while also removing spurious self-accelerations such as those described above. 

Detweiler-Whiting prescriptions provide an organizing principle for many results on the motion of self-interacting bodies \cite{Poisson2011, Harte_2015, pound}. Variants are now known to hold not only for point particles, but also for arbitrary extended objects -- in both Klein-Gordon \cite{Harte_2008} and Maxwell \cite{HarteEM} theories, and at least through linear order in general relativity \cite{harteeffectivemetric}. Similar results hold also in spacetimes with different numbers of dimensions \cite{Harte:2016fru, Harte:2018iim}. In all of these cases, the same effective fields describe not only the evolution of a body's linear momentum, but also its angular momentum. Restricting at least to translational motion at monopolar order, even the second-order gravitational self-force is known to admit a Detweiler-Whiting prescription \cite{pound1, pound3, pound4}. 

These prescriptions are not merely summaries of existing self-force results. They can also be derived directly, and doing so has proven to be a particularly efficient route to understanding motion \cite{Harte_2015}. However, most such derivations have been restricted to linear theories. We remove that restriction here, presenting a framework which both generalizes and simplifies the derivation of Detweiler-Whiting prescriptions. Although one of our primary motivations is to understand motion in general relativity -- where detailed knowledge of the gravitational self-force is needed to optimally interpret future gravitational-wave observations \cite{pound} -- we largely focus on a model problem: objects coupled to nonlinear scalar fields. 

We construct a large class of field transformations which leave the exact laws of motion form-invariant. Practical approximations then arise by selecting those transformations which remove enough small-scale variation from the physical fields to justify low-order multipolar expansions. In point-particle limits, these transformations remove infinities and can be described as regularizations. Our framework thus allows point-particle regularizations to be derived from first principles. 

Interesting field transformations for extended bodies, including the Newtonian transformation $\phiN \to \phiNhat$, are necessarily nonlocal. This can be understood by noting that, e.g., the center of mass $z$ of an extended body is a nonlocal functional of its configuration. Changes in that centroid cannot therefore depend only on physical fields at $z$. If we demand that all laws of motion nevertheless be written in terms of local operators at $z$, the only option is to apply those operators to nonlocal effective fields. 

As an example, consider an extended body with scalar charge density $\rho$ which is coupled to a massless Klein-Gordon field $\phi$ in a four-dimensional spacetime with fixed metric $g_{ab}$. Laws of motion are then known to be simple in terms of the Detweiler-Whiting ``R field''\footnote{We use the notation ${\hat \phi}(x; \rho, \phi, g_{ab}]$ to denote that $\hat{\phi}$ can depend in an ordinary way on the point $x$ and also functionally on $\rho$, $\phi$, and $g_{ab}$.} \cite{Detweiler:2002mi, Harte_2008, Poisson2011}
\be
   {\hat \phi}_{\rm DW} (x; \rho, \phi, g_{ab}] \equiv \phi(x) - \int \! \rmd V' \rho(x')
   G_{\rm DW}(x,x'; g_{ab}] ,
    \label{hatphi00}
\ee
where $G_{\rm DW}$ denotes the ``S-type'' Detweiler-Whiting Green function and $\rmd V' = \sqrt{-g(x')} \rmd^4 x'$. Being a Green function, $\nabla^a \nabla_a G_{\rm DW} =\delta$ so $\nabla^a \nabla_a \hat{\phi}_{\rm DW} = 0$. Furthermore, $G_{\rm DW}(x,x'; g_{ab}]$ is symmetric under the interchange of $x$ and $x'$. It also vanishes wherever those points are timelike-separated.  In flat spacetime, $G_{\rm DW} = \frac{1}{2} (G_{\rm advanced} + G_{\rm retarded})$. Regardless, while $\hat{\phi}_{\rm DW}$ is useful in Klein-Gordon theory, more complicated effective fields are needed in more complicated theories. 

\textit{Effective stress-energy tensors and effective scalar fields --} Now consider the motion of an extended body coupled to a (possibly nonlinear) scalar $\phi$. The action which governs this system,
\begin{equation}
  S = \Sm [\chi_A, \phi, g_{ab}] + \Sf [\phi,g_{ab}],
\label{eq:action0}
\end{equation}
is assumed to split into a matter component $\Sm$ and a field component $\Sf$, the latter of which does not depend on the short-ranged matter fields $\chi_A$ which compose the body of interest. Both $\Sm$ and $\Sf$ are assumed to be preserved by diffeomorphisms: If $\sigma^*$ denotes the pullback associated with any diffeomorphism $\sigma$,
\be
\Sm[ \sigma^* \chi_A, \sigma^* \phi, \sigma^* g_{ab}] =
\Sm[  \chi_A,  \phi, g_{ab}],
\label{eq:covariant1}
\ee
and similarly for $\Sf$. 

Using $\Lie_\zeta$ to denote the Lie derivative with respect to a vector field $\zeta^a$, linearized diffeomorphism-invariance implies that
\be
0 = \int \! \rmd^4 x \left[
  \frac{\delta \Sm }{\delta \chi_A} \Lie_\zeta \chi_A +
  \frac{\delta \Sm }{\delta \phi} \Lie_\zeta \phi +
  \frac{\delta \Sm}{\delta g_{ab}} \Lie_\zeta g_{ab} 
  \right]
  \label{dSmatt}
\ee
for all $\zeta^a$. If we define the charge density $\rho$ and the matter component $T^{ab}_{\rm matt}$ of the stress-energy tensor via
\be
    \rho \equiv -\frac{1}{\sqrt{-g}} \frac{ \delta \Sm }{ \delta \phi}, \qquad  T^{ab}_{\rm matt} \equiv \frac{2}{\sqrt{-g}} \frac{ \delta  \Sm }{ \delta g_{ab} },
    \label{rhoTDef}
\ee
varying over all $\zeta^a$ with compact support while also applying the equations of motion $\delta \Sm/\delta \chi_A = 0$ shows that
\be
    \nabla_b T^{ab}_{\rm matt} = - \rho \nabla^a \phi.
    \label{eq:cons2}
\ee
A similar argument applied to the full action implies that the total stress-energy tensor $T^{ab} \equiv (2 / \! \sqrt{-g} ) \delta  S / \delta g_{ab}$ is conserved in the sense that $\nabla_b T^{ab} = 0$ \cite[Appendix E]{Wald}. Lastly, the field equation $\delta S/\delta \phi = 0$ can be written as ${\cal E}(x;\phi,g_{ab}] = \rho$, where $\sqrt{-g} {\cal E} \equiv \delta \Sf / \delta \phi.$ 

Our goal now is to find mappings which preserve appropriate laws of motion. The laws we consider below derive from the conservation equation \eqref{eq:cons2}, so this can be accomplished by showing that there exist effective or ``dressed''  stress-energy tensors $\hat{T}^{ab}_{\rm matt}$, charge densities $\hat{\rho}$, and fields ${\hat \phi}$ which obey the effective conservation equation
\begin{equation}
	\nabla_b \hat{T}^{ab}_{\rm matt} = - \hat{\rho} \nabla^a \hat{\phi}. 
	\label{stressconsHat}
\end{equation}
Any law of motion applicable to the original ``bare'' variables then retains its form in terms of the dressed variables. The need to dress (or finitely renormalize) $T^{ab}_{\rm matt}$ can be understood physically by noting that, e.g., an object's self-field has inertia which must affect its motion. Although charge densities are not affected by the Detweiler-Whiting field transformation used in simple theories, there are more complicated systems in which screening and other effects can arise.

It was shown in \cite{Harte_2008, HarteEM, harteeffectivemetric, Harte_2015, Harte:2016fru, Harte:2018iim, harteEffT} that the Detweiler-Whiting prescription \eqref{hatphi00} preserves laws of motion because forces appear, inside a body, in action-reaction pairs which cancel up to terms that can be absorbed into redefinitions of the linear and angular momenta. Such cancellations depend, in part, on the symmetry of the Detweiler-Whiting Green function $G_{\rm DW}$. Now, rearranging \eqref{hatphi00} to write the physical field $\phi_{\rm DW}$ in terms of a Detweiler-Whiting effective field $\hat{\phi}$,
\begin{equation}
	G_{\rm DW} (x,x';g_{ab}] = \frac{1}{\sqrt{-g'}} \frac{ \delta \phi_{\rm DW} (x;\rho, \hat{\phi}, g_{ab}] }{ \delta \rho(x') } .
\end{equation}
The symmetry of $G_{\rm DW}$ can thus be guaranteed by supposing that $\phi_{\rm DW}$ is the functional derivative with respect to $\rho$ of some ``generating functional'' $W[\rho, \hat{\phi}, g_{ab}]$.

One of our primary insights is to suppose that this remains true even when going beyond the original Detweiler-Whiting prescription: All field maps we consider are assumed to derive from a generating functional $W[\rho, \hat{\phi}, g_{ab}]$ which satisfies
\begin{equation}
	\phi = \frac{1}{\sqrt{-g}} \frac{ \delta W [\rho, \hat{\phi},g_{ab}]}{ \delta \rho } .
	\label{WDef}
\end{equation}
This implicitly defines $\hat{\phi}$ in terms of $\phi$, $\rho$, and $g_{ab}$. For any such mapping, $n$-point functions obtained by repeatedly differentiating $\phi$ with respect to $\rho$ are automatically symmetric in all $n$ points.

We now show that \textit{any diffeomorphism-invariant generating functional defines effective variables which satisfy the usual laws of motion}. First note that just as the diffeomorphism invariance \eqref{eq:covariant1} of $\Sm$ implies \eqref{dSmatt} and \eqref{eq:cons2}, the diffeomorphism invariance of $W$ implies that
\begin{equation}
	\rho \nabla^a \phi = \frac{1}{\sqrt{-g}} \frac{\delta W }{ \delta \hat{\phi}} \nabla^a \hat{\phi} - \nabla_b \left( \frac{2}{\sqrt{-g}} \frac{ \delta W }{ \delta g_{ab} } - \rho \phi g^{ab} \right) .
\end{equation}
Identifying $\hat{\rho}$ and ${\hat T}^{ab}_{\rm matt}$ with
\begin{equation}
	\hat{\rho} \equiv \frac{1}{\sqrt{-g}} \frac{ \delta W}{ \delta \hat{\phi}} , \qquad {\hat T}^{ab}_{\rm matt} \equiv T^{ab}_{\rm matt} -
	\frac{2}{\sqrt{-g}} \frac{\delta W}{\delta g_{ab}} + \rho \phi g^{ab} ,
	\label{HattedVars1}
\end{equation}
substitution into the bare conservation equation \eqref{eq:cons2} then recovers the effective conservation equation \eqref{stressconsHat}. Different effective variables can thus be found simply by specifying different diffeomorphism-invariant functionals. Physically, the corresponding mappings convert some of the long-ranged interactions mediated by $\phi$ into short-ranged internal stresses.

The simplest relevant field map is the identity, which is generated by
\begin{equation}
	W_0[ \rho, \hat{\phi}, g_{ab} ] \equiv \int \! \rmd V \rho \hat{\phi}.
\end{equation}
This leaves $\phi$, $\rho$, and $T^{ab}_{\rm matt}$ as-is. More complicated functionals are, of course, necessary in self-force problems. For example, the Detweiler-Whiting mapping \eqref{hatphi00} follows from
\begin{equation}
    W = W_0[ \rho, \hat{\phi}, g_{ab} ] + \frac{1}{2} \int \! \rmd V \rmd V' \rho \rho' G_{\rm DW}(x,x';g_{ab}].
    \label{DWmap}
\end{equation}
Although $\hat{\rho} = \rho$ here, $\hat{T}^{ab}_{\rm matt} \neq T^{ab}_{\rm matt}$. Even in this well-studied context, a non-perturbative expression for $\hat{T}^{ab}_{\rm matt}$ had previously been derived only for static systems \cite{Harte:2016fru}. Eq. \eqref{HattedVars1} generalizes that result to fully dynamical scenarios and also to much more general field mappings. Unlike Detweiler-Whiting prescriptions for the evolution of a body's linear and angular momenta, this has no known analog in Newtonian gravity.

Our primary interest here is in field mappings which are more general than \eqref{DWmap}. Although there are many possibilities which preserve the exact laws of motion, the majority of these are not useful; certain selection principles must be applied to the space of diffeomorphism-invariant generating functionals. One such principle is that there should be no remapping at all in the test-body limit where self-fields are negligible. This suggests that $W = W_0 + \mathscr{O}(\rho^2)$.

A more stringent selection principle demands that low-order multipole expansions for the force and torque be good approximations for the actual force and torque in a wide range of physical systems. That occurs when a mapping removes from $\phi$ any lengthscales which are comparable to or smaller than a body's size. One way to enforce this is to demand that $W$ be chosen such that on shell, $\hat{\phi}$ is a solution to the homogeneous field equation 
\begin{equation}
    \mathcal{E} (x; \hat{\phi}, g_{ab} ] = 0.
    \label{zeroSource}
\end{equation}
This implies that the $n$-point functions which can be used to specify $W$ [cf. \eqref{eq:Wexpansion} below] must satisfy certain partial differential equations.

Our final selection principle states that although physically-interesting mappings are necessarily nonlocal, they should not be \textit{too} nonlocal. The $\hat{\rho}$ and $\hat{T}^{ab}_{\rm matt}$ associated with a compact body should not depend on anything far away from that body, nor on anything in the distant past or future. We now restrict attention only to \textit{finite} bodies, namely those for which $T^{ab}_{\rm matt}$ and $\rho$ both vanish outside of a spatially-compact worldtube $\mathcal{B}$. We also restrict consideration to some neighborhood $\mathcal{N}$ of $\mathcal{B}$. Physically-reasonable mappings should then obey the following quasilocality conditions:
\begin{enumerate}
	\item There exists a proper subset $\hat{\mathcal{B}}$ of $\mathcal{N}$, which is a neighborhood of $\mathcal{B}$, such that $\hat{\rho}$ and $\hat{T}^{ab}_{\rm matt}$ both vanish on $\mathcal{N} \setminus \hat{\mathcal{B}}$.
        
	\item Given any $x \in \mathcal{N}$, the effective variables $(\hat{\phi}, \hat{\rho}, \hat{T}^{ab}_{\rm matt} )$ at $x$ depend on the physical variables $( \phi, g_{ab}, \rho, T^{ab}_{\rm matt} )$ only within a finite neighborhood of that point.
\end{enumerate}  
Condition 1 represents a kind of spatial quasilocality. It demands that if a body's physical worldtube $\mathcal{B}$ is spatially compact, so is its effective worldtube $\hat{\mathcal{B}}$. Combining this with condition 2 enforces a temporal quasilocality in which the effective variables cannot depend on anything in the distant past or future. Note that condition 1 is nontrivial even in linear theories, where $\hat{T}^{ab}_{\rm matt}$ might be expected to include a contribution from the infinitely-extended ``self-self'' component of the field's stress-energy tensor. While this expectation can be tenable in certain perturbative contexts \cite{GrallaHarteWald}, it is more generally unphysical to expect that, e.g., an object's inertia depends on distant fields.

\textit{Laws of motion --} Thus far, we have largely identified a body's ``laws of motion'' with the conservation equation \eqref{eq:cons2} or its effective counterpart \eqref{stressconsHat}. However, laws of motion more commonly refer to evolution equations for, e.g., a body's center of mass. It is convenient to discuss such concepts in terms of a ``generalized momentum'' which is constructed by integrating $T^{ab}_{\rm matt}$ over an appropriate hypersurface \cite{HarteSyms, Harte_2015}. Given a physically-reasonable generating functional $W$, we adapt this concept to define an effective (or dressed) generalized momentum via
\begin{equation}
    \hat{\calP}_\xi (s) \equiv \int_{\Sigma_s} \hat{T}_{\rm matt}^{ab}\,  \xi_a \rmd S_b.
    \label{eq:genmom}
\end{equation}	
This assumes that a spacelike foliation of $\hat{\mathcal{B}}$, with leaves $\Sigma_s$, has been chosen, as well as a set of relevant vector fields $\xi^a$. In flat spacetime, all such vector fields should be Killing. More generally, we can employ generalized Killing fields \cite{HarteSyms, Harte_2015}. Regardless, $\hat{\calP}_\xi = \hat{p}_a \xi^a + \frac{1}{2} \hat{S}^{ab} \nabla_a \xi_b$ encodes both the ``ordinary'' linear momentum $\hat{p}_a$ as well as the angular momentum $\hat{S}^{ab} = \hat{S}^{[ab]}$, both of which can be recovered by varying over all $\xi^a$. If $W=W_0$, the resulting bare momenta  coincide with the momenta originally introduced by Dixon \cite{mpdd, dixon1, Dix79}. 
    
Dixon showed that if his momenta are obtained from a spatially-compact stress-energy tensor which is appropriately conserved, they evolve according to the Mathisson-Papapetrou-Dixon (MPD) equations \cite{dixon1, Dix79}. Although he did not consider bodies with scalar charge, the relevant generalizations are straightforward \cite{Harte_2008, Harte_2015}. Our effective stress-energy $\hat{T}^{ab}_{\rm matt}$ is spatially compact and satisfies an effective conservation equation which has the same form as the bare conservation equation, so effective MPD equations follow immediately: If $z(s)$ parametrizes the worldline with respect to which the momenta are defined, $\hat{F}_a(s)$ denotes an appropriate force, and $\hat{N}^{ab} = \hat{N}^{[ab]}(s)$ an appropriate torque, 
\begin{subequations}
    \label{MPD}
    \begin{align}
        \frac{\rm D}{\rmd s} \hat{p}_a &= \frac{1}{2} R_{bcda} \hat{S}^{bc} \dot{z}^d + \hat{F}_a,
	\\
	\frac{\rm D}{\rmd s} \hat{S}^{ab} &= 2 \hat{p}^{[a}
        \dot{z}^{b]} + \hat{N}^{ab}.
    \end{align}
\end{subequations}
These are the ordinary MPD equations when $W=W_0$. Different choices for covariantly-constructed generating functionals in which $\hat{T}^{ab}_{\rm matt}$ remains spatially compact instead result in \textit{different momenta which still satisfy ``MPD-like'' laws of motion}. Forces and torques are determined by varying $\xi^a$ in the generalized force
\begin{align}
    \frac{\rmd}{\rmd s} \hat{\calP}_\xi &= \hat{F}_a \xi^a + \frac{1}{2} \hat{N}^{ab} \nabla_a \xi_b, 
    \nonumber
    \\ 
    &= \int_{\Sigma_s} \left( \frac{1}{2} \hat{T}_{\rm matt}^{ab} \Lie_\xi
    g_{ab} - \hat{\rho} \Lie_\xi \hat{\phi} \right) t^a \rmd S_a, 
    \label{PdotBare}
\end{align}
where $t^a$ is any vector field satisfying $t^a \nabla_a s =1$. This results in exact integrals for $\hat{F}_a$ and $\hat{N}^{ab}$ which are analogous to the exact Newtonian acceleration \eqref{accelN}. Contributions here from $\Lie_\xi g_{ab}$ or $\Lie_\xi \hat{\phi}$ are interpreted as gravitational or scalar effects, respectively  (the former of which vanish in flat spacetime). The bare versions of the gravitational force and torque may be found explicitly in, e.g., \cite{dixon1, Dix79}. 

If there are significant short lengthscales in $\hat{\phi}$ or $g_{ab}$, the integral expressions for the force and torque cannot be usefully expanded in terms of multipole moments. Physically, this is interpreted as meaning that $\hat{F}_a$ and $\hat{N}^{ab}$ depend sensitively on a body's internal structure. However, if $\hat{\phi}$ and $g_{ab}$ both vary slowly throughout $\Sigma_s$, there is only a weak dependence on internal structure, multipolar expansions are practical, and the first nontrivial terms are  \cite{mpdd, dixon1, Dix79, Harte_2015}
\begin{equation}
    \hat{F}_a = -\frac{1}{6} \hat{J}^{bcde} \nabla_a R_{bcde} - \hat{q} \nabla_a \hat{\phi}, \quad \hat{N}_{ab} = \frac{4}{3} \hat{J}^{cdf}{}_{[a} R_{b]fcd}.
\end{equation}
Here, $\hat{q}$ denotes the total charge associated with $\hat{\rho}$ and $\hat{J}^{abcd} = \hat{J}^{[ab]cd} = \hat{J}^{ab[cd]} = \hat{J}^{cdab}$ the quadrupole moment associated with $\hat{T}^{ab}_{\rm matt}$. Selecting a ``good'' $W$ thus amounts to identifying bulk degrees of freedom $\hat{p}_a$ and $\hat{S}^{ab}$ whose evolution equations are largely universal. We expect that in most cases of physical interest where this is possible -- meaning that short lengthscales can be removed from $\phi$ -- it can be accomplished by enforcing the source-free condition \eqref{zeroSource}, or perhaps a slight generalization thereof.

By themselves, the MPD equations evolve $\hat{p}_a$ and $\hat{S}^{ab}$, but not $z$. Nevertheless, one would typically use these equations to evolve a representative worldline: a ``center of mass.'' One way to do so is to impose the Tulczyjew-Dixon spin-supplementary condition $\hat{p}_a \hat{S}^{ab} = 0$. This implicitly selects a particular worldline \cite{mpdd, Dix79}, at least when $\hat{p}_a \hat{p}^a \neq 0$ \cite{HarteOancea}. It also provides a relation between $\hat{p}^a$ and $\dot{z}^a$ which allows $z$ to be evolved along with the momenta. This procedure is well-known for the bare variables \cite{EhlersRudolph, Dix79, Costa2015, Harte_2015} and it works identically for any effective variables. Known existence and uniqueness results for the center of mass \cite{CM1,CM2} also remain valid when applied to $\hat{T}^{ab}_{\rm matt}$ rather than $T^{ab}_{\rm matt}$. 
	
\textit{Weak nonlinearity expansion --} Our results thus far reduce the problem of motion to the construction of a diffeomorphism-invariant generating functional $W$ which satisfies the aforementioned selection criteria. Given a particular theory, it is not necessarily obvious that such a functional exists, and even when it does exist, it may be difficult to construct. 

Rather than working in full generality in a nonlinear theory, we now restrict attention to systems which are only weakly nonlinear. The generating functional can then be expanded in the functional Taylor (or Volterra) series
\begin{align}\label{eq:Wexpansion}
		W = W_0[\rho, \hat{\phi}, g_{ab}] +  \sum_{n=2}^\infty \frac{1}{n} \int \rmd V_1 \cdots \rmd V_n \, \rho_1 \cdots \rho_n 
		\nonumber
		\\
		~{} \times G_{(n)}  (x_1, \ldots , x_n; {\hat \phi}, g_{ab}],
\end{align}
where the $n$-point functions $G_{(n)}$ are fully symmetric in their arguments and transform covariantly under diffeomorphisms.

Truncating at $n=3$, the quasilocality conditions above can be shown to be satisfied by assuming that the 2- and 3-point functions have the following properties:
\begin{enumerate}
    \item[A.] $G_{(2)}(x_1,x_2;\hat{\phi},g_{ab}] = 0$ wherever $x_1$ and $x_2$ are timelike-separated. 

    \item[B.] $G_{(3)}(x_1, x_2, x_3; \hat{\phi}, g_{ab}] = 0$ wherever two or more of the three possible pairs formed from $x_1$, $x_2$, and $x_3$ are timelike-separated.

    \item[C.] If $\hat{\phi}$ or $g_{ab}$ are varied, the corresponding variations $\delta G_{(2)}$ and $\delta G_{(3)}$ can depend on $\delta \hat{\phi}$ and $\delta g_{ab}$ at most within the convex hulls\footnote{We define the convex hull $\mathscr{C}(R)$ of a region $R$ to be the smallest set containing $R$ such that, for any two points in $\mathscr{C}(R)$, there exists a geodesic segment which connects those points and which is contained entirely in $\mathscr{C}(R)$. 
    } $\mathscr{C}(\{x_1,x_2\})$ or $\mathscr{C}(\{x_1,x_2,x_3\})$.
\end{enumerate}
Combining \eqref{HattedVars1} and \eqref{eq:Wexpansion}, properties A and C imply that the contribution of $G_{(2)}$ to $\hat{\rho}$ or to $\hat{T}^{ab}_{\rm m}$ is confined to the set
\begin{equation}
    \bigcup \left\{ \left.  \mathcal{C} \big( \{ x_1,x_2\} \big) \, \right| \mbox{non-timelike $x_1,x_2 \in \mathcal{B}$} \right\}.
\end{equation}
This is a spatially-compact worldtube, as demanded by quasilocality condition 1. Properties B and C instead imply that the contribution of $G_{(3)}$ lies in the union of all convex hulls of the triples $x_1,x_2,x_3 \in \mathcal{B}$ in which no more than one pair in any triple is timelike-separated. This too is spatially compact. Similar reasoning may also be used to verify our second quasilocality condition.

A useful approximate generating functional can thus be obtained by finding symmetric and covariant 2- and 3-point functions with properties A--C above. These functions should also satisfy field equations which guarantee \eqref{zeroSource} through an appropriate order. Such constraints generalize the Detweiler-Whiting prescription for weakly-nonlinear systems. 

\textit{An example theory --} In order to illustrate our formalism, consider the field action $\Sf = -(2 \lambda^2)^{-1} \int \rmd V \, | \nabla e^{\lambda \phi} |^2$, where $\lambda >0$ is fixed. This implies the nonlinear field equation
\begin{align}
	\nabla^a \nabla_a \phi + \lambda |\nabla \phi|^2   = \rho e^{-2\lambda \phi} .
    \label{nonlinearField}
\end{align}
Motion in similar theories has previously been discussed using effective field theory \cite{GalleyScalar1, GalleyScalar2}.

For our purposes, an effective field map can be found by noting that when $\rho = 0$, the variable $\psi \equiv e^{\lambda \phi}$ satisfies the linear equation $\nabla^a \nabla_a \psi = 0$. A modified Detweiler-Whiting prescription can thus be applied to $\psi$ and then transformed into a prescription for $\phi$. Again using $G_{\rm DW}$ to denote an S-type Detweiler-Whiting Green function satisfying $\nabla^a \nabla_a G_{\rm DW} = \delta$, the resulting effective field is
\begin{align}
	\hat{\phi}(x;\rho, \phi,g_{ab}] = \frac{1}{\lambda} \ln \bigg[ e^{\lambda \phi} - \lambda \int \rmd V' \rho' e^{-\lambda \phi'} 
	 G_{\rm DW} \bigg].
	\label{phiHatEx}
\end{align}
This satisfies the homogeneous field equation $\nabla^a \nabla_a \hat{\phi} + \lambda | \nabla \hat{\phi}|^2 = 0$. The simplicity of this theory also allows all $n$-point functions to be written in terms of the 2-point function $G_{(2)} (x_1,x_2;\hat{\phi},g_{ab}] = e^{-\lambda (\hat{\phi}_1 + \hat{\phi}_2) } G_{\rm DW} (x_1, x_2; g_{ab}]$. For example, 
\begin{align}
    G_{(3)} (x_1, x_2, x_3; \hat{\phi},g_{ab}] = - \frac{3}{2} \lambda \sym_{x_1,x_2,x_3} \Big( G_{(2)} (x_1, x_2; \hat{\phi},g_{ab}] 
    \nonumber
    \\
    ~ \times ~ G_{(2)}(x_1, x_3; \hat{\phi},g_{cd}] \Big),
\end{align}
where ``$\sym$'' denotes a symmetrization over the listed variables. 

Although $G_{\rm DW}$ does not depend on the field, $G_{(2)}$ does. Charges are therefore dressed. This effect can be computed by noting that the charge density $\rho/\lambda \psi$ which is associated with $\psi$ is not dressed when transforming to $\hat{\psi} \equiv e^{\lambda \hat{\phi}}$. The generalized scalar force density associated with $\hat{\psi}$ is thus proportional to $(\rho/\lambda \psi) \Lie_\xi \hat{\psi} =  \rho e^{\lambda (\hat{\phi} - \phi)} \Lie_\xi \hat{\phi}$. Therefore,
\begin{equation}
	\hat{\rho} = \rho e^{\lambda ( \hat{\phi} - \phi)}.
\end{equation}
If $\rho$ has spatially-compact support, so does $\hat{\rho}$. 

At leading order in a multipole expansion, the momentum of an object with dressed charge $\hat{q}$ evolves via ${\rm D}\hat{p}_a/\rmd s = - \hat{q} \nabla_a \hat{\phi}$. All that remains in order to obtain a more explicit equation of motion is the implementation of a precise point-particle limit and the computation of an associated $\phi$.

\textit{Gravitational theories --} Our discussion thus far has assumed that the metric is fixed. If it is instead dynamical, as it would be in a gravitational self-force problem, we would like a map which relates the physical metric $g_{ab}$ to a more slowly varying effective metric $\hat{g}_{ab}$. Now ignoring any non-gravitational fields, one way to accomplish this is to consider a diffeomorphism-invariant generating functional $W[\mathscr{T}_{cd}, \hat{g}^{ab}]$, where $\mathscr{T}_{ab} \equiv ( \delta^c_a \delta^d_b - \tfrac{1}{2} g_{ab} g^{cd} ) T_{cd}$ denotes the trace-reversed stress-energy tensor. Making the identifications
\begin{align}
    g^{ab} = \frac{1}{\sqrt{-g}} \frac{ \delta W}{\delta \mathscr{T}_{ab}}, \qquad \hat{T}_{ab} = \frac{1}{\sqrt{ - \hat{g} }} \frac{ \delta W}{ \delta \hat{g}^{ab} },
\end{align}
diffeomorphism invariance implies that if $T_{ab}$ is conserved in the sense that $g^{ab} \nabla_a T_{bc} = 0$, the dressed stress-energy is conserved with respect to the dressed metric: $\hat{g}^{ab} \hat{\nabla}_a \hat{T}_{bc} = 0$. If $\hat{T}_{ab}$ has spatially-compact support, it follows that the (purely gravitational) MPD equations hold for momenta defined using $\hat{g}_{ab}$ and $\hat{T}_{ab}$. Furthermore, if the effective metric varies sufficiently slowly, its motion will be approximately geodesic with respect to $\hat{g}_{ab}$. Detailed applications are, however, left for later work.

\input{shortFin2.bbl}
%\bibliography{Ref.bib}

\end{document}

%% file: shortFin2.bbl
%apsrev4-2.bst 2019-01-14 (MD) hand-edited version of apsrev4-1.bst
%Control: key (0)
%Control: author (8) initials jnrlst
%Control: editor formatted (1) identically to author
%Control: production of article title (0) allowed
%Control: page (0) single
%Control: year (1) truncated
%Control: production of eprint (0) enabled
%